\documentclass[]{spie}  

 \usepackage{siunitx}
\usepackage{amsmath,amsfonts,amssymb}
\usepackage{graphicx}
\usepackage{subcaption}
\usepackage{float}
\usepackage[colorlinks=true, allcolors=blue]{hyperref}

\title{Exploring the Capabilities of Astrophotonics for the Precise Alignment of Segmented Telescopes}

\author[a]{Maria Cuevas*}
\author[b]{Aditya R. Sengupta}
\author[c]{Vincent Chambouleyron}
\author[b]{Rebecca Jensen-Clem}
\author[b]{Daren Dillon}
\author[d]{Sylvain Cetre}
\author[b]{Ma\"{i}ssa Salama}
\author[e]{Caleb Dobias}
\author[e]{Tara Crowe}
\author[e]{Stephen S. Eikenberry}
\author[e]{Rodrigo Amezcua-Correa}
\author[e]{Stephanos Yerolatsitis}

\affil[a]{Department of Astronomy, Columbia University, New York, NY, USA}
\affil[b]{Department of Astronomy and Astrophysics, University of California, Santa Cruz, CA 95064, USA}
\affil[c]{Laboratoire d'Astrophysique de Marseille}
\affil[d]{Wakea Consulting, Durham, UK}
\affil[e]{CREOL, The College of Optics \& Photonics, University of Central Florida, Orlando, FL 32816, USA}

\authorinfo{Further author information: (Send correspondence to *M.C.)\\ *M.C.: E-mail: mc5300@columbia.edu}
\begin{document} 
\maketitle

\begin{abstract}

 The next generation of large telescopes for direct imaging of exoplanets will require segmented primary mirrors. Over both long and short timescales, these telescopes experience segment misalignments which degrade the final science image. Adaptive optics (AO) systems can be used to correct these aberrations in real time. AO systems require wavefront sensors (WFSs) that measure the phase of the incoming light in order to reconstruct optical aberrations. However, most WFSs used for sensing atmospheric turbulence cannot correctly detect aberrations induced by misalignments in segmented telescopes, as they show poor sensitivity to phase discontinuities. We investigate the potential of photonic lanterns (PLs), which are waveguides that allow for the low-loss transmission from multi-mode to multiple single-mode optical signals, for sensing segment misalignments at the focal plane. We assess the ability of PLs to measure piston offsets in segmented mirrors through both simulations and laboratory experiments. We simulate the photonic lantern and demonstrate linear reconstruction on segment pistons. Further, we train a neural network to reconstruct aberrations outside of the linear regime. We experimentally validate reconstruction of segment piston offsets on the Miniature Infrared SEAL (muirSEAL) testbed, which includes a segmented deformable mirror, a PSF imaging branch, and a PL. This work demonstrates the potential of the PL as a compact WFS for future space- and ground-based segmented-mirror telescopes.

\end{abstract}


\section{INTRODUCTION}
Segmented telescopes require extremely precise alignment and stability to maximize the sensitivity of exoplanet imaging instruments. Aberrations induced by piston, tip, and tilt offsets of individual segments of the primary mirror can significantly degrade image quality. Traditional slope-based pupil-plane wavefront sensors (WFSs) are not able to detect these aberrations, and they are additionally not able to detect non-common-path aberrations (NCPAs) that affect only the final science image. Focal-plane wavefront sensors may be able to simultaneously detect segment aberrations and NCPAs, making them valuable for high-contrast imaging.

Closed-loop control of segment aberrations using the Zernike WFS has been demonstrated on sky at the Keck II telescope\cite{Salama24}. However, the Zernike WFS is not able to detect all NCPAs as it is not the science imager. This motivates the need for a focal-plane WFS that not only helps with segment phasing, but can also detect NCPAs.

This work explores the use of photonic lanterns (PLs) \cite{Birks2015,Norris20} as focal-plane WFSs for segmented-aperture telescopes. A photonic lantern guides multi-mode input light into an array of single-mode outputs, with the output intensity distribution encoding information about the incoming wavefront \cite{Lin2022}. 

PLs are known to be sensitive to aberrations due to pupil plane phase discontinuities such as the low-wind effect \cite{Wei24}, making them a potentially suitable choice for correcting aberrations due to misalignments in the primary mirrors of segmented telescopes. Additionally, PLs can be used to build science instruments like spectrographs by taking advantage of single-mode specific devices like fiber Bragg gratings \cite{Gatkine2019,Ellis2020}. This makes them a versatile component for high-contrast imaging.

We characterize the lantern's sensitivity to segment-level aberrations through both simulation and laboratory experiments. In this work, we develop simulations to quantify the lantern response to segment piston errors, and conduct laboratory experiments of segment piston reconstruction using a near-infrared testbed with a photonic lantern and segmented deformable mirror. We find that photonic lanterns offer a promising solution for segment phasing in future space- and ground-based high-contrast imaging systems. 

The remainder of this paper is structured as follows. Section 2 discusses simulations of aberrated segmented mirrors and the use of a PL to detect and correct these aberrations. In Section 3 we show laboratory experiments of the photonic lantern and assess the performance of real-life PLs for linear reconstruction. Finally, we conclude in section 4. 

\section{Simulation Work}
\subsection{Simulation Parameters}

We simulate a hexagonal segmented aperture using the \texttt{make\_hexagonal\_segmented} class within the \textit{hcipy} \cite{Por18} Python package. To simulate piston aberrations on the segmented mirror, we use the \texttt{SegmentedDeformableMirror} class. We propagate the wavefront after the segmented deformable mirror to the focal plane and through the photonic lantern.

\begin{table}[h!]
\centering
\begin{tabular}{|l|c|}
\hline
\multicolumn{2}{|l|}{\textbf{Simulation Parameters}} \\
\hline
Number of pixels, pupil plane     & 1025 px \\
Wavelength       & 1.55$\mu$m  \\
f-number at PL entrance            & 8 \\
Segment flat-to-flat distance & 146.276 px\\
Number of DM segments   & 36 \\
PL cladding refractive index  & 1.4504\\
PL core refractive index & 1.46076\\
PL cladding diameter & 37$\mu$m \\
PL core diameter & 4.4$\mu$m\\
PL core offsets from center for inner ring & 7.4$\mu$m\\
PL core offsets from center for outer ring & 14.8$\mu$m\\
PL length& 60mm\\
PL tapering factor & 8\\

\hline
\end{tabular}
\caption{Description of simulation input parameters.}
\label{tab:sim_params}
\end{table}

We simulate a standard (non-mode-selective) 19-port PL using the \texttt{lightbeam} Python package\cite{lightbeam}. We use the same PL design parameters as in Sengupta et al. 2024\cite{Sengupta24}. We re-state these parameters and list the relevant \textit{hcipy} simulation parameters in Table~\ref{tab:sim_params}.


\begin{figure}

    \centering
     
    \begin{subfigure}[b]{0.6\textwidth}
        \centering
        \includegraphics[width=\textwidth]{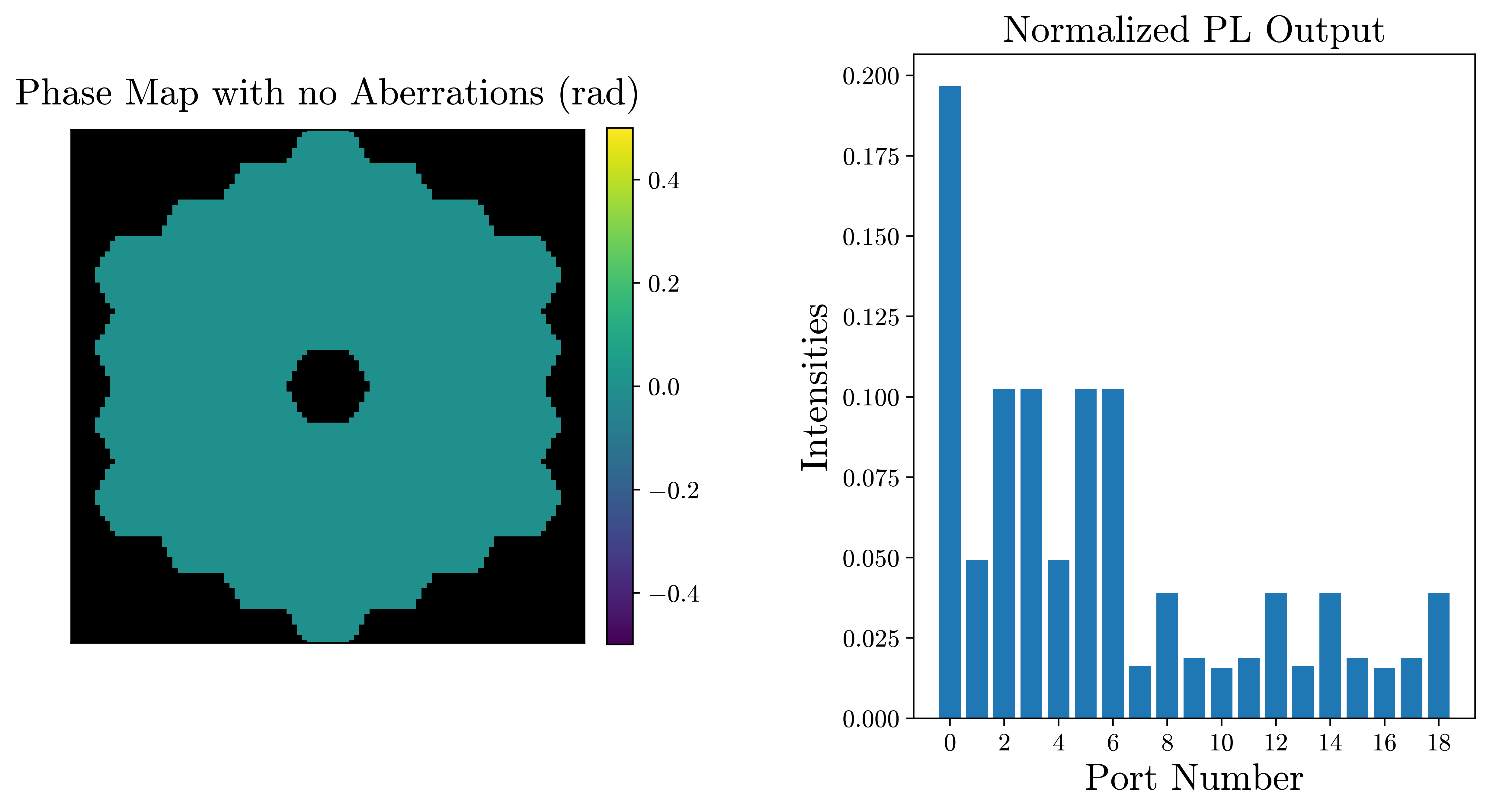}
        \caption{} 
        \label{fig:lantern_response1}
    \end{subfigure}
    \begin{subfigure}[b]{0.6\textwidth}
        \centering
        \includegraphics[width=\textwidth]{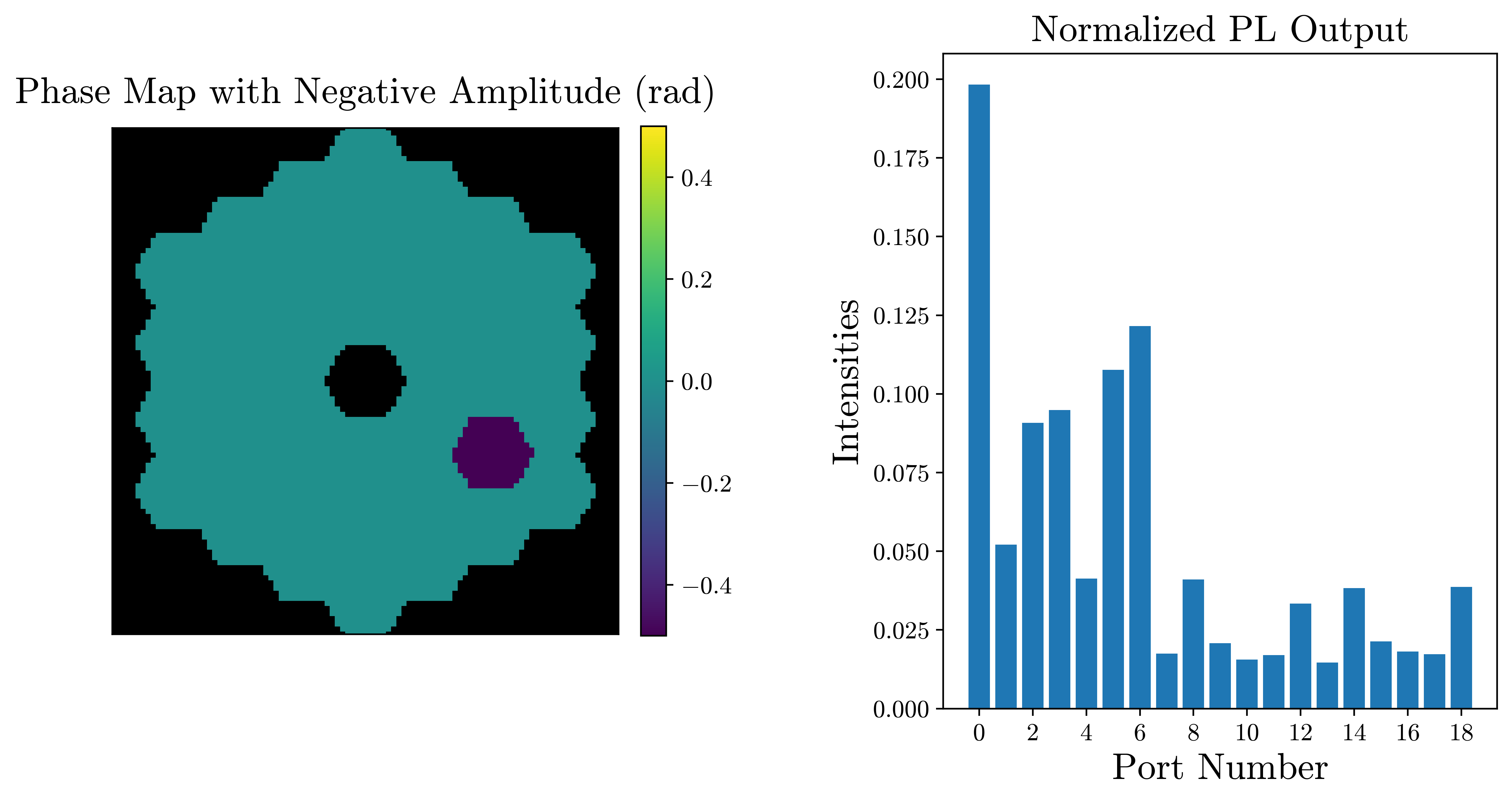}
        \caption{} 
        \label{fig:lantern_response2}
    \end{subfigure}

    \begin{subfigure}{0.6\textwidth}
        \centering
        \includegraphics[width=\textwidth]{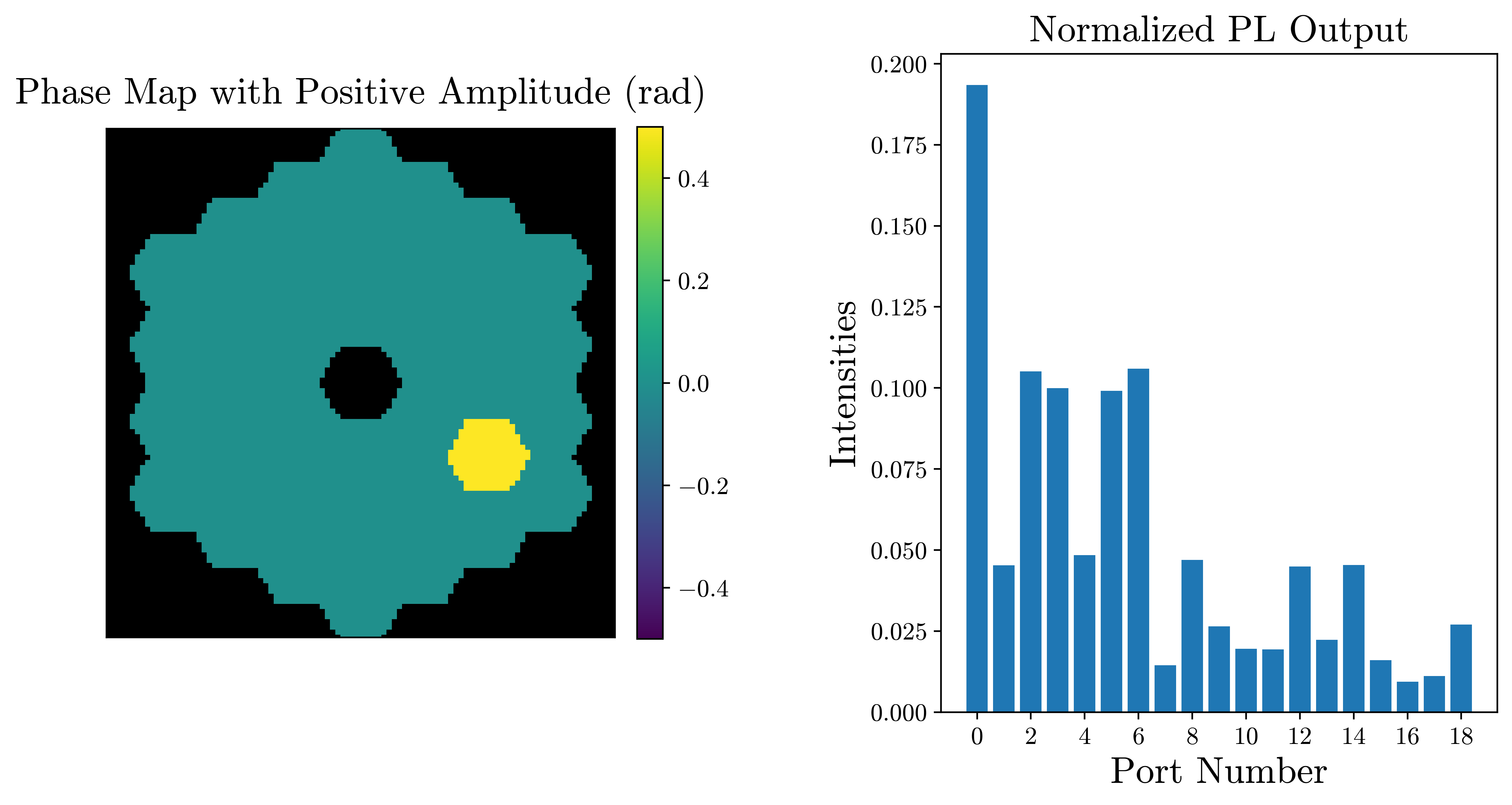}
        \caption{} 
        \label{fig:lantern_response3}
    \end{subfigure}

    \caption{The lantern is sensitive to positive and negative piston aberrations. 
    Panel (a) shows the lantern's response to flat segment, (b) showcases the lantern's intensities in response to a negative aberration whereas (c) shows a distinct lantern response to a positive aberration of equal magnitude.}
    \label{fig:combined_lantern_response}
\end{figure}

Figure~\ref{fig:combined_lantern_response} shows that the PL response encodes which segments show piston offsets, as well as their amplitude.

\subsection{Linear Reconstruction}
\label{sec:simlin}

We employ the standard WFS technique of modeling the photonic lantern as a linear system with respect to segment pokes \cite{Lin2022,Lin2023}. Although optical propagation is nonlinear in intensity, for small aberrations, there exists a linear regime that allows us to perform linear reconstruction. To achieve this, we construct a push-pull interaction matrix by applying small positive and negative segment piston errors ($10^{-10}$ meters) to the segmented DM and recording the normalized output at each of the 19 PL ports in response to these aberrations. The resulting matrix is then inverted to form a command matrix. The command matrix multiplied by the normalized and flat-subtracted PL intensity provides the linear reconstruction of the injected phase.


\begin{figure}
    \centering
    \includegraphics[width=0.5\linewidth]{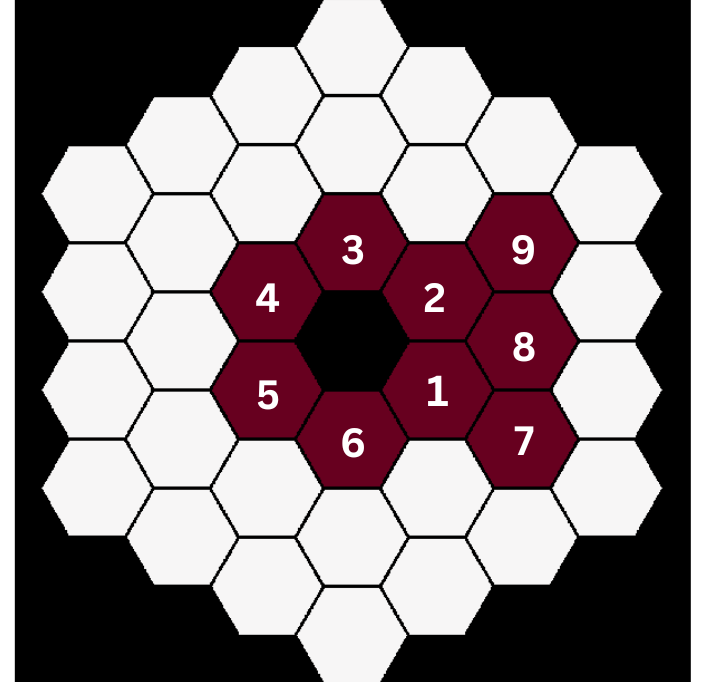}
    \caption{The locations of the nine segments for which we reconstruct piston offsets in this work.}
    \label{fig:segnums}
\end{figure}

\begin{figure}
    \centering
    \includegraphics[width=0.6\textwidth]{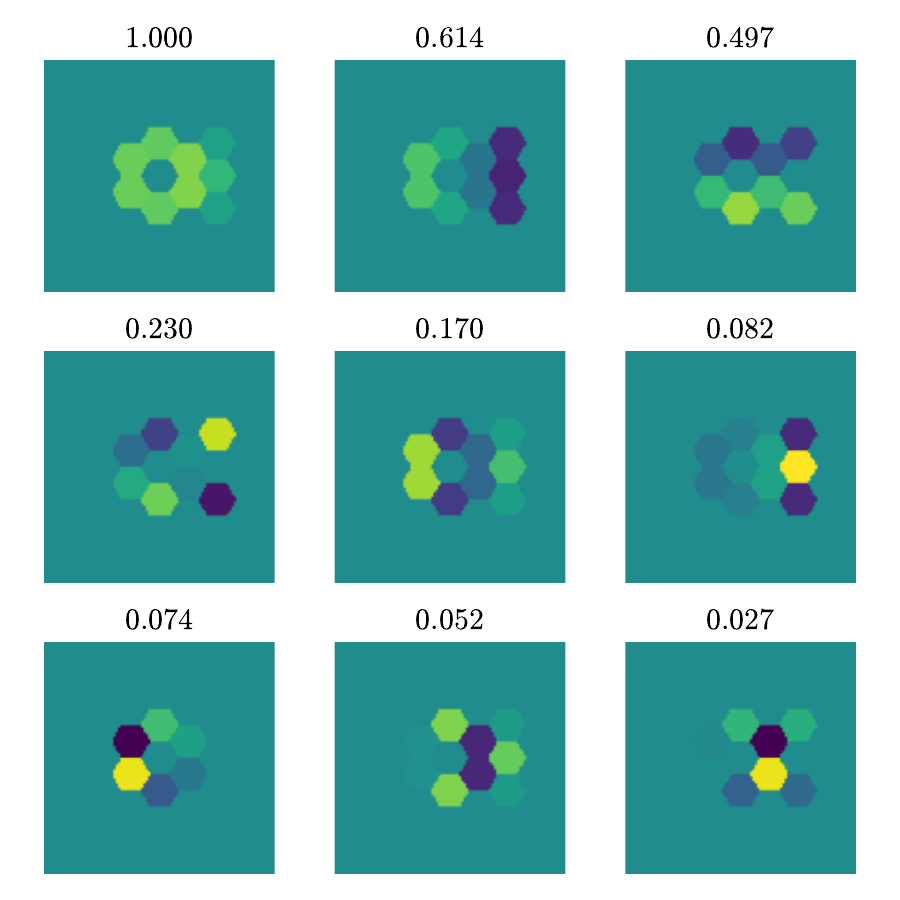}
    \caption{Eigenmodes and singular values associated with the simulated interaction matrix.}
    \label{fig:eigenmodes}
\end{figure}

Figure~\ref{fig:eigenmodes} shows the eigenmodes in segment piston space associated with the interaction matrix. The singular values for each mode are shown above. We note $x$-axis symmetries that are not present across the $y$-axis; for instance, segments 1 and 2 (as specified in Figure~\ref{fig:segnums}) show the same response in most eigenmodes. This may be due to artificial symmetries from the PL simulation grid lining up precisely with the \textit{hcipy} grid, and we expect this not to carry over to laboratory testing, as small system aberrations and rotations would break this symmetry.

\begin{figure}
    \centering
    \includegraphics[width=0.85\linewidth]{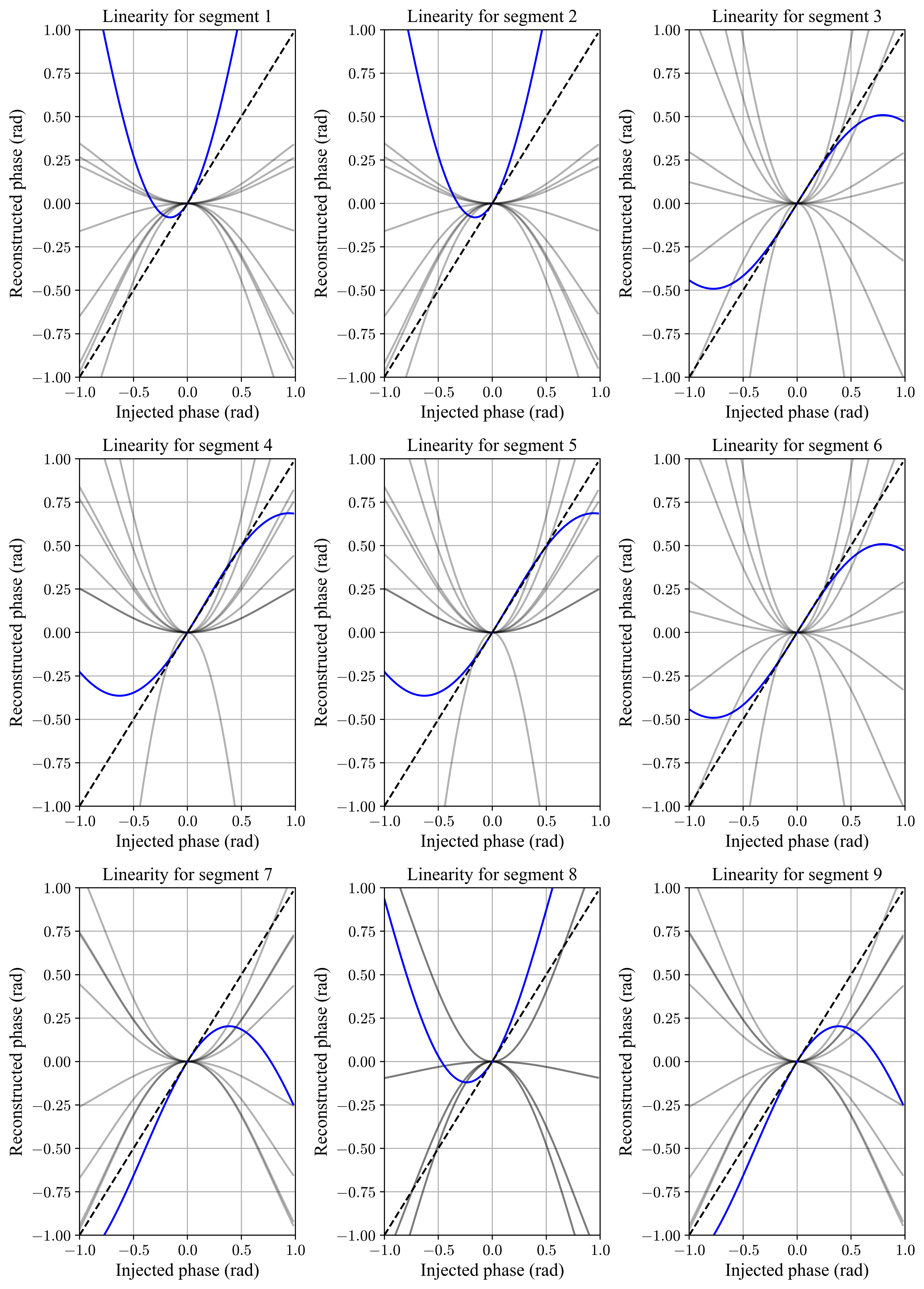}
    \caption{Linearity curves when reconstructing segments using an interaction matrix.}
    \label{fig:linsim}
\end{figure}

Figure~\ref{fig:linsim} shows linearity curves for the nine segments shown in Figure~\ref{fig:segnums}. For each segment, we inject amplitudes from -1 to 1 radians and reconstruct the signal using the command matrix. We plot blue curves for the injected segment and gray for the remaining segments to showcase crosstalk. These plots show that we can perform linear reconstruction for small piston aberrations. For some individual segments we observe a linear range of $\sim 0.5$ radians. Beyond this regime we observe crosstalk (substantial reconstructed error on segments that were not actually poked) meaning that for large amplitudes, or for multi-segment aberrations that are not well expressed as linear combinations of the eigenmodes in Figure~\ref{fig:eigenmodes}, we would not be able to use a linear reconstructor. For segments 1, 2, and 8, we observe quick overturning and small linear ranges. These may behave differently from the remaining segments as a result of the previously-noted artificial symmetry issue; segments 1 and 2 previously showed symmetry within the eigenmodes, and segment 8 is the only one that we consider which lies on the $x$-axis of the pupil. The small linear ranges in some cases, together with the crosstalk, motivate the use of machine learning to build a non-linear reconstructor.

\subsection{Neural Network Reconstruction}

We extend our earlier simulations to minimize the crosstalk between segments. We implemented the multi-layer neural network first demonstrated by Norris \textit{et al.}\cite{Norris20} and with the same architecture as used by Sengupta et al. 2024.\cite{Sengupta24}. This model was created using the python library \textit{pytorch}. It uses two hidden layers with 2000 and 100 neurons respectively, and uses a ReLU activation function. We use a training set of 45,000 measurements, a mean squared error cost function, and the Adam optimizer with a learning rate of 0.001. The dataset for the model was created by randomly applying piston offsets in the interval of -1 to 1 radians to segments 1 through 9 (Figure ~\ref{fig:segnums}).

\begin{figure}
    \centering
    \includegraphics[width=0.85\linewidth]{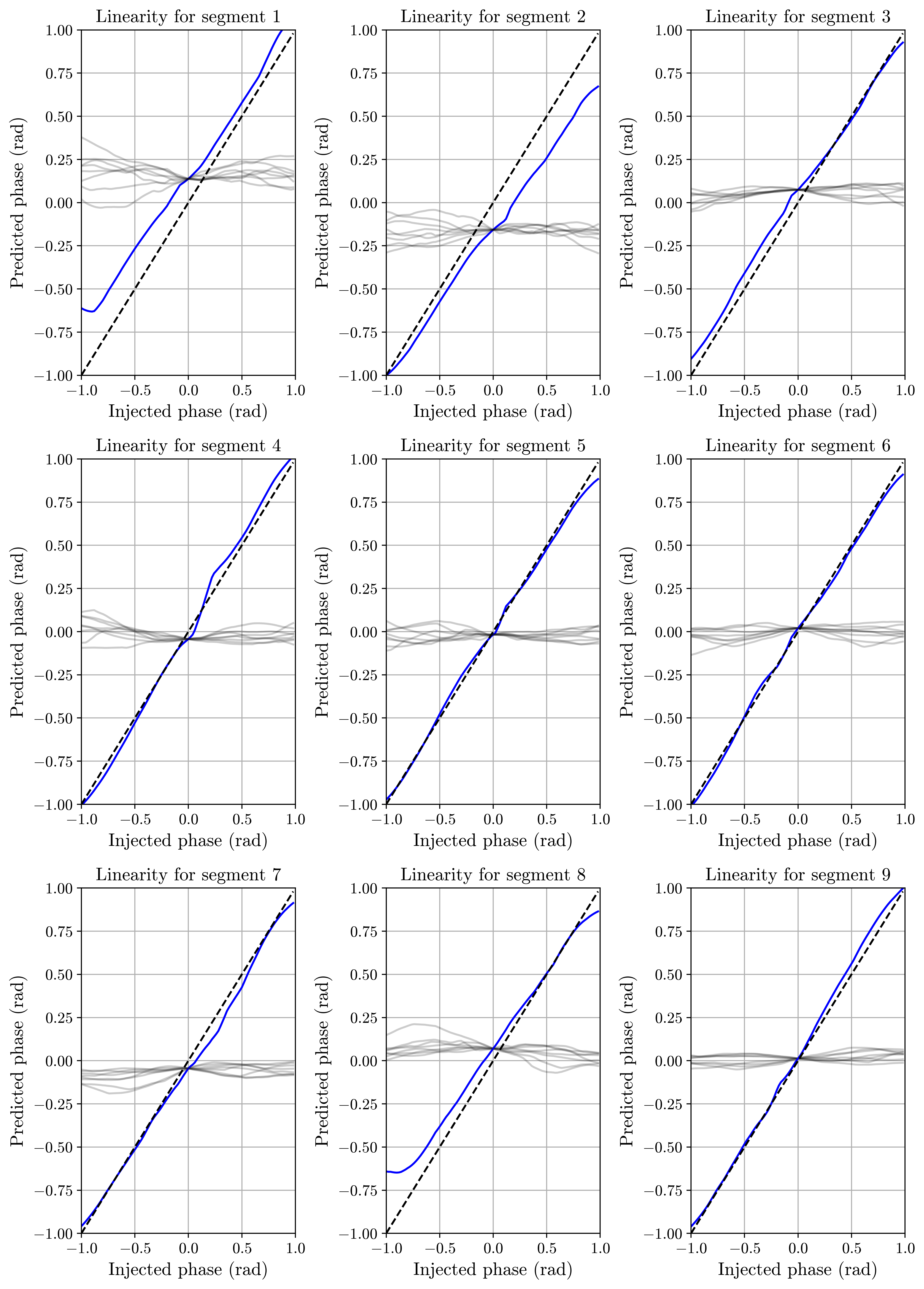}
    \caption{Dynamic ranges when reconstructing segments using a neural network model.}
    \label{fig:nn}
\end{figure}

We made response curves, shown in Figure ~\ref{fig:nn} to assess how well our model reconstructed aberrations along the range of -1 to 1 radians. We limited our tests to this range in order to match the typical amplitudes of segment piston aberrations \cite{Salama24}; however, future work may be able to extend this. We note that for multiple segments the model is not able to accurately predict when the segment has no piston offsets. Adding in a direct linear component, i.e. a matrix connecting the input and output with no hidden layers or activation function, has proven useful to resolve a similar issue for a neural network reconstructor for a pyramid WFS\cite{Landman2024}. This would allow the neural network to provide corrections to deviations from the linear reconstruction, which may enable it to match the performance for small aberrations shown by the linear method.

\subsection{Closed-Loop Control}

\begin{figure}
    \centering
    \includegraphics[width=0.85\textwidth]{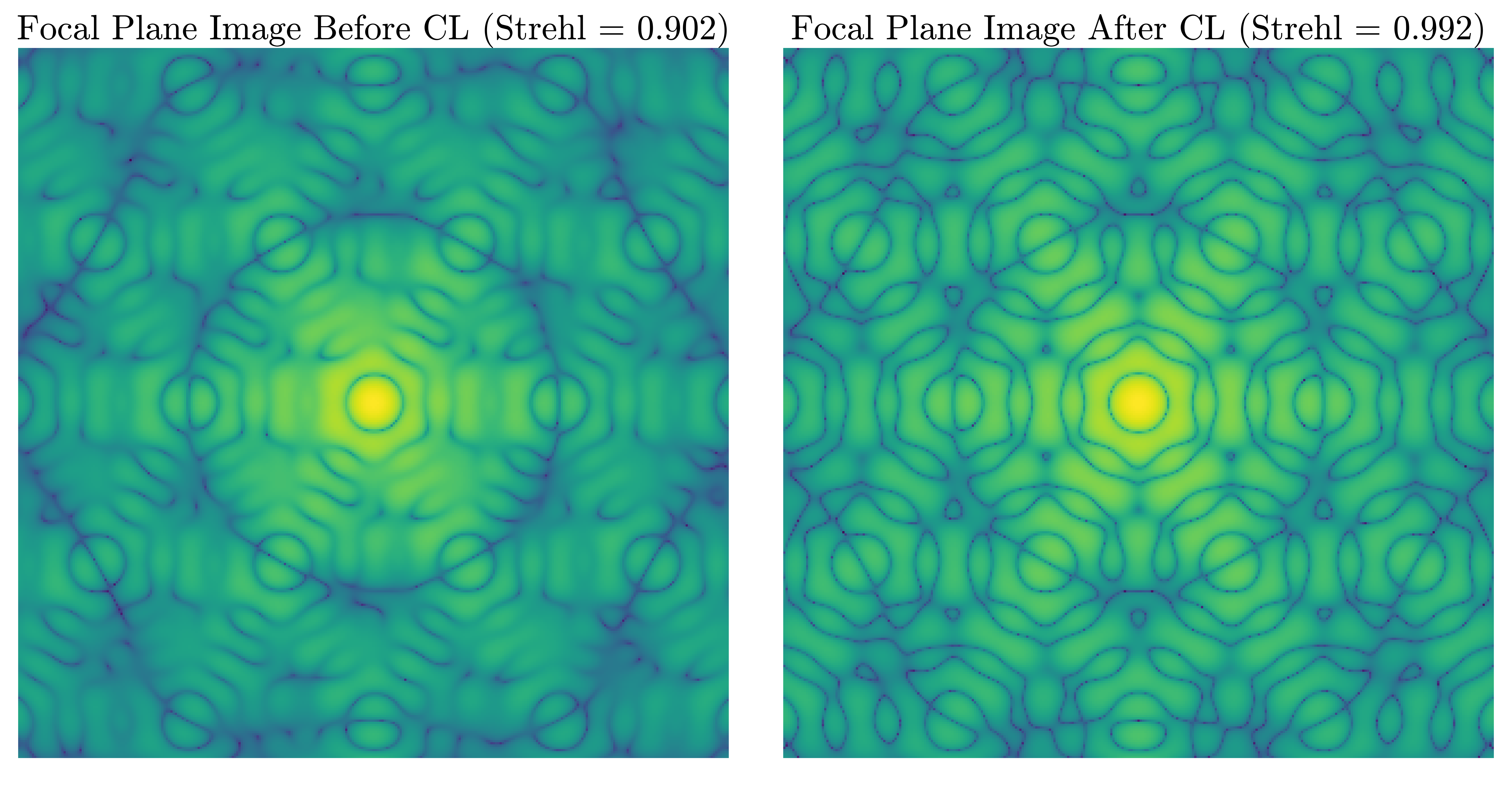}
    \caption{Point spread function through a simulated segmented mirror before and after closed-loop control. Prior to applying any corrections the image had a Strehl ratio of 0.902 and the RMS of the piston errors was \SI{82.61}{\nano\meter}. After closed-loop control the Strehl ratio increased  to 0.992 and the RMS piston error decreased to \SI{0.273}{\nano\meter}. }
    \label{fig:psf_before_after}
    
\end{figure}

We demonstrate closed-loop control of piston aberrations using the PL. We apply an initial aberration to the chosen nine segments. We select these aberrations such that they are small enough to be within the linear regime of all the segments. We then record the corresponding PL intensities, and reconstruct the wavefront using the linear reconstructor. The result is then applied to the actuators of the deformable mirror. We run the loop for 50 iterations and measure the Strehl ratio of the resulting PSF. 
Figure~\ref{fig:psf_before_after} shows the Strehl ratio improving after the loop is closed. The initial RMS of the piston errors on the DM was \SI{82.61}{\nano\meter}, which decreased to \SI{0.273}{\nano\meter} after closed-loop control.

\section{Laboratory experiments}
\label{sec:sections}

We perform laboratory experiments using the muirSEAL testbed (described in detail by Sengupta \textit{et al.} in these proceedings\cite{Sengupta25}) as part of the SEAL high-contrast imaging testbed in the UC Santa Cruz Laboratory for Adaptive Optics\cite{JensenClem21SEAL,SEAL2}. The testbed consists of an IrisAO segmented deformable mirror and a 19-port photonic lantern. 

We use the muirSEAL testbed to apply piston aberrations to the segments in the beam and measure the lantern intensities. We apply aberrations in the range $-1$ rad to 1 rad (approximately $\pm 0.25 \mu$m). We take the full PL image as our wavefront sensing signal, rather than carrying out aperture photometry to obtain per-port intensities. We then use these measurements to make an interaction matrix over only the six innermost segments (using a poke amplitude of 0.1 radians) as outlined in Section ~\ref{sec:simlin} and assess linearity.

\begin{figure}
    \centering
    \includegraphics[width=0.85\linewidth]{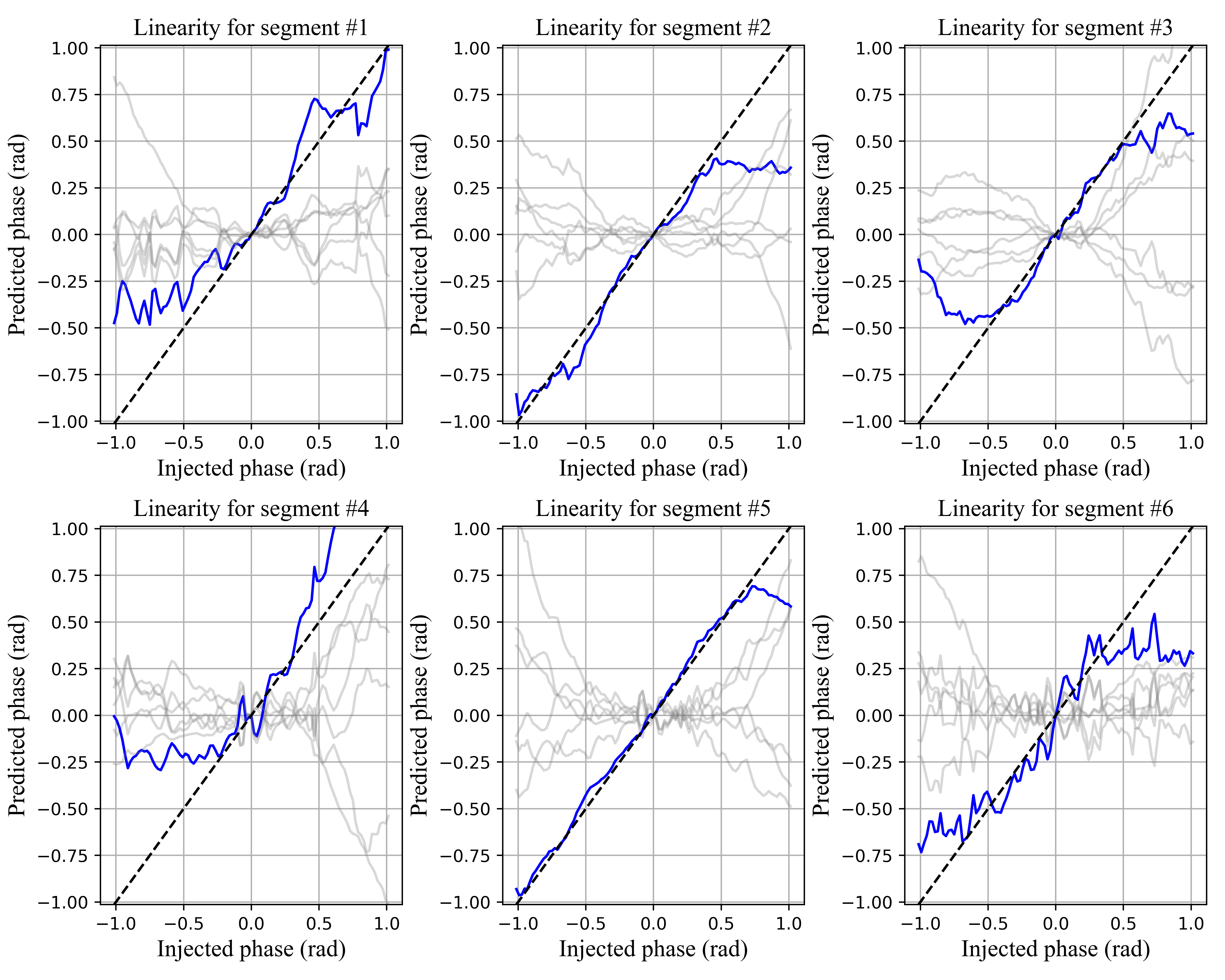}
    \caption{Linearity curves from Miniature IR SEAL (muirSEAL) in the UCSC Lab for Adaptive Optics.}
    \label{fig:linbench}
\end{figure}

In figure ~\ref{fig:linbench} we observe wider linear ranges than our simulations and minimal cross-talk at small aberrations. We observe a linear range of 0.5 radians (123.35 nm) which is consistent with the linear ranges seen in simulation. These are also comparable to table 2 of Salama \textit{et al.}\cite{Salama24} where a total RMS of 128 nm is reported after running a closed-loop test with the Zernike wavefront sensor. This suggest that our PL system may achieve similar closed-loop performance. 

We note that, unlike the simulation, there are no segments which show quick overturning or significantly smaller linear ranges than the others. As discussed previously, this suggests that these features were a result of artificial symmetries within the simulation. However, we also note significant variation within the linearity curves. For example, in the linearity curve for segment 4, the dynamic range appears to be wide, but the local behavior at small aberrations (around 0 rad) shows an immediate inversion, i.e. a positive poke being read as negative and vice versa. This suggests that the PL does encode wavefront information about this segment, but the measurements may have uncharacterized noise. Further analysis of the PL's noise properties and the signal-to-noise ratio of bench measurements is required.





\section{Conclusion}

We present results from PL wavefront reconstruction methods both in simulation and in the laboratory. We simulate a segmented primary mirror and a photonic lantern and demonstrate linear reconstruction. We then extend our work to neural networks to address cross-talk between segments. Lastly, we demonstrate linear reconstruction in a laboratory setting using the muirSEAL testbed. Our results show the ability of photonic lanterns to perform segment phasing. 

Future work will include (i) extending our laboratory work to nonlinear reconstructors and closed-loop control (ii) adjusting our simulations to be more accurate to the laboratory setup and (iii) looking at noise propagation in the lantern. We expect that these steps will help to more fully characterize this system's dynamic range, sensitivity, and the number of controllable segments.

This work on photonic lanterns for segment phasing will be critical for high-contrast imaging on the next generation of segmented telescopes. 

\acknowledgements

MC thanks Deana Tanguay and Enrico Ramirez-Ruiz from the Lamat Institute for their unwavering support. This work was catalyzed during the Lamat REU program supported by NSF grant 2150255.
\bibliography{report} 
\bibliographystyle{spiebib} 

\end{document}